\documentclass{pnastwo}

\usepackage{amssymb,amsfonts,amsmath}
\usepackage{caption}
\usepackage{graphicx,bm}
\usepackage{epsfig}
\usepackage{color}
\usepackage{epsfig,dsfont,amssymb,amsmath,amsthm,amsfonts,amsbsy,mathrsfs}

\usepackage{ulem}

\contributor{Submitted to Proceedings
of the National Academy of Sciences of the United States of America}
\url{www.pnas.org/cgi/doi/10.1073/pnas.0709640104}
\copyrightyear{2008}
\issuedate{Issue Date}
\volume{Volume}
\issuenumber{Issue Number}

\begin{document}



\title{Scaling description of the yielding transition in soft amorphous solids at zero temperature}





 \author{Jie Lin\affil{1}{Center for Soft Matter Research, Department of Physics, New York University, New York, NY 10003},
  Edan Lerner\affil{1}{}, Alberto Rosso\affil{2}{Laboratoire de Physique Th{\'e}orique et Mod{\`e}les  Statistiques (UMR CNRS 8626), Universit\'e de Paris-Sud, Orsay Cedex, France } ,
 \and Matthieu Wyart\affil{1}{}}

\contributor{Submitted to Proceedings of the National Academy of Sciences
of the United States of America}


\maketitle
\begin{article}

\begin{abstract}
Yield stress materials flow if a sufficiently large shear stress is applied. Although such materials are ubiquitous and relevant for industry,
there is no accepted microscopic description of how they yield, even in the simplest situations where temperature is negligible and  where flow inhomogeneities such as shear bands or fractures are absent. 
Here we propose a scaling description of the yielding transition in amorphous solids made of soft particles at zero temperature. Our description makes  a connection between  the Herschel-Bulkley exponent characterizing the singularity of the flow curve near the yield stress $\Sigma_c$, the extension and duration of the avalanches of plasticity observed at threshold, and the density $P(x)$ of soft spots, or shear transformation zones, as  a function of the stress increment $x$ beyond which they yield. We argue   that the critical exponents of the yielding transition can be expressed in terms of three independent exponents $\theta$, $d_f$ and $z$, characterizing respectively the density of soft spots, the fractal dimension of the avalanches, and their duration.  Our description shares some similarity with the depinning transition that occurs when  an elastic manifold is driven through a random potential, but also presents some striking differences.  We test our arguments in an elasto-plastic model, an automaton model similar to those used in depinning, but with a different interaction kernel, and find satisfying agreement with our predictions both in two and three dimensions.
\end{abstract}


\section{Significance}
Yield stress solids flow if a sufficiently large shear stress is applied. Although such materials are ubiquitous and relevant for industry, there is no accepted microscopic description of how they yield. Here we propose a scaling description of the yielding transition which relates  the flow curve, the statistics of the avalanches of plasticity observed at threshold, and the density of local zones that are about to yield.  Our description shares some similarity with the depinning transition that occurs when an elastic manifold is driven through a random potential, but also presents some striking differences. Numerical simulations on a simple elasto-plastic model find good agreement with our predictions.

 \section{Introduction}
Many solids will flow and behave as fluids if a sufficiently large shear stress is applied. In crystals, plasticity is governed by the motion of dislocations 
\cite{Weiss2001,Zaiser2006}. In amorphous solids there is no order, and conserved defects cannot be  defined. However, as noticed by Argon \cite{argon}, plasticity consists of elementary events localized in space, called shear transformations, where a few particles rearrange.  This observation supports that there are special locations in the sample, called shear transformation zones or STZ's \cite{langer98}, where the system lies close to an elastic instability.  Several theoretical approaches of plasticity, such as STZ theory \cite{langer98}  
or Soft Glassy Rheology (SGR) \cite{sollich1998} assume that such zones relax independently, or are coupled to each other via an effective temperature.  However, at zero temperature and small applied strain rate $\dot\gamma$, computer experiments \cite{maloney2010,caroli,lerner,salerno,Maloney2004,Gimbert2013} and very recent experiments \cite{Amon2012,crassous} indicate that local rearrangements are not independent: plasticity occurs  via avalanches in which many shear transformations are involved, forming  elongated structures where plasticity localizes. If conditions are such that  flow is homogeneous (as can occur for example in foams or emulsions),  one finds that  the flow curves are singular at small strain rate and follow a Herschel-Bulkley law $\Sigma-\Sigma_c \sim \dot\gamma^{1/\beta}$ \cite{hohler2005rheology,barnes01}.  These features are qualitatively reproduced by elastoplastic models \cite{Talamali2011,barret,kirsten,Picard2005,Lin2014} where space is discretized. In these models, a site that yields plastically affects the stress in its surroundings via some interaction kernel ${\cal G}(r)$, argued to  decay as a power-law of distance and to display a four-fold symmetry \cite{Picard2004},  as supported by observations \cite{Maloney2006,Chattoraj2013,Schall2007}. This perturbation can trigger novel plastic events and lead to avalanches. However, even within this picture, the relationship between the avalanche dynamics and the singularity of the flow curves remains debated \cite{caroli,Bocquet2009}. 

\begin{figure}[htb!]
\label{f11}
\centering\includegraphics[width=.989\linewidth]{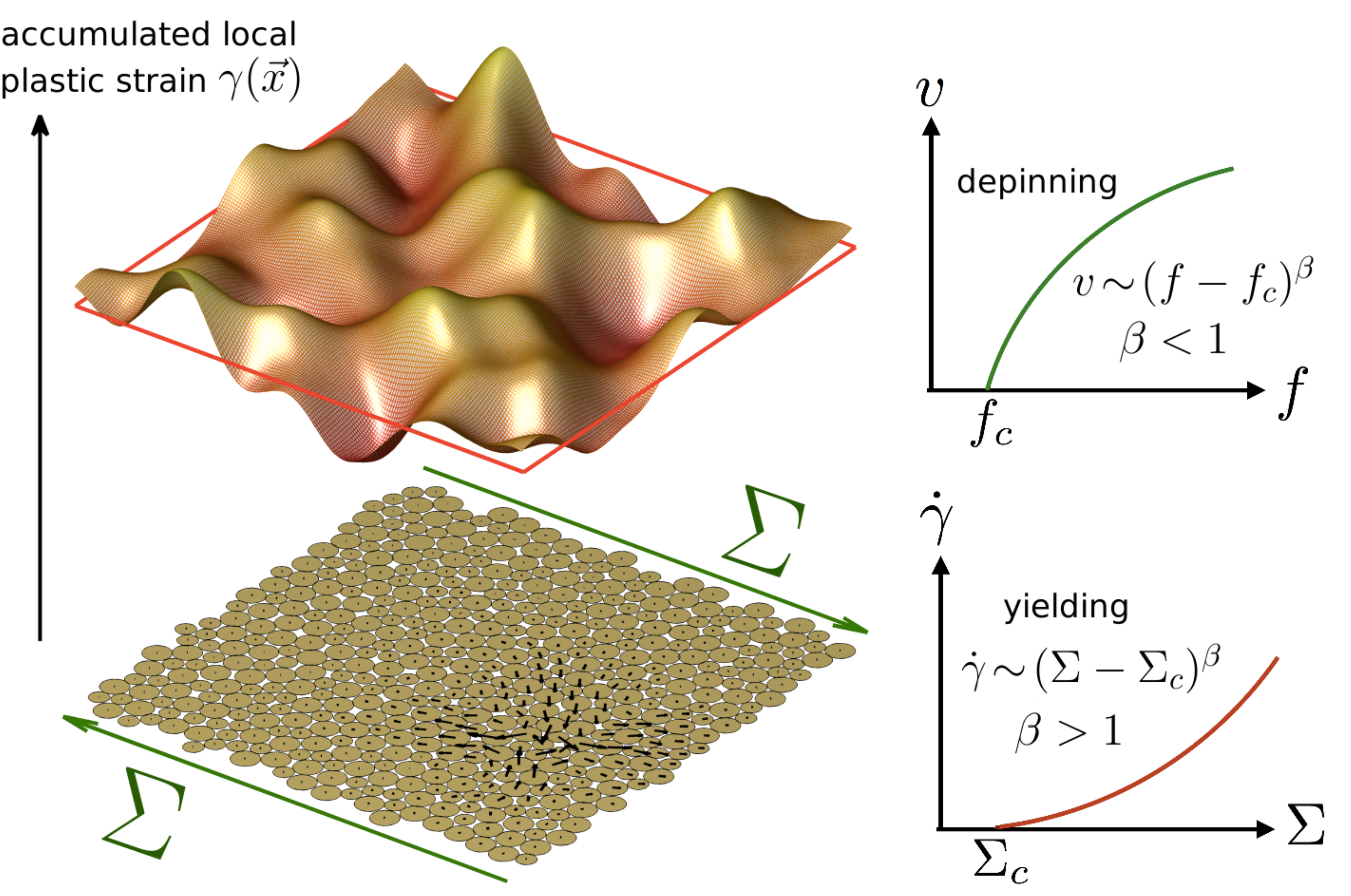}
\caption{\small{Left: Analogy between the yielding transition of a $d$-dimensional amorphous solid and the depinning transition of an elastic interface of $d$ dimensions in a space of $d+1$ dimensions, illustrated here for $d=2$. The height of the interface is the accumulated local plastic strain generated by local plastic rearrangements; one example of the latter appears  in the bottom. Right: the strain rate-stress (velocity-force) curves for the yielding (depinning) transition with $\beta>1$ ($\beta<1$) in yielding (depinning) transition.}}
\end{figure}

It is tempting to seek progress by building a comparison between the yielding transition and the much better understood depinning transition that occurs when an elastic interface  of dimension $d$ is driven in a $d+1$ random environment \cite{Zaiser2006,Fisher1998}.  The role that transverse displacements play in depinning corresponds to the local accumulated plastic strain $\gamma({\vec x})$ and the total plastic strain $\gamma$ can be identified with the center of mass of the interface,  as illustrated in Fig.(\ref{f11}).  Both phenomena display very similar properties: near the depinning threshold force $F_c$, the velocity $v$ vanishes non-analytically $v\sim (F-F_c)^\beta$ and the interplay between disorder and elasticity at threshold leads to broadly distributed avalanches corresponding to jerky motions of the interface. Much more is known about the depinning transition: it is a dynamical critical point characterized by two independent exponents related to avalanche extension and  duration \cite{Fisher1998,Kardar1998}. These exponents have been computed perturbatively with the functional renormalisation group \cite{Nattermann1992, Narayan1993,  Chauve2001}  and evaluated numerically with high precision \cite{Ferrero2013}.The comparison between these two phenomena has led to the proposition that the yielding transition is in the universality class of mean-field depinning \cite{Dahmen2013,dahmen2011simple}. However,  experiments find a reological exponent  $\beta>1$ against the $\beta \le 1$ predicted for elastic depinning, and numerical simulations display intriguing finite size effects that differ from depinning \cite{salerno,Maloney2004,Edan2010,Salerno2013,lemaitre2007plastic}.
 
Formally, elastoplastic models  are very similar to  automaton models known to capture well the depinning transition \cite{barret}, the key difference relies in the interaction kernel ${\cal G}$, long-ranged and of variable sign for elastoplastic models while is essentially a laplacian for depinning with short-range elasticity. We have recentely shown \cite{Lin2014} that in presence of long-ranged with variable sign interactions the distribution of shear transformations at a distance $x$ from instability, $P(x)$, is singular with $P(x)\sim x^\theta$, unlike in depinning for which $\theta=0$. As we shall recall this singularity naturally explains the finite size effects observed in simulations. In this letter we argue that once this key difference with depinning is taken into account, the analogy between these two phenomena  is fruitful, and leads to a complete scaling description of the yielding transition. In particular we find that the Herschel-Bulkley exponent is related to avalanche extension and duration via Eq.(\ref{beta}) and that the avalanche statistics can be expressed in term of three independent exponents: $\theta$, $d_f$ and $z$, characterizing respectively the density of shear transformations, the fractal dimension of the avalanches, and their duration.


\section{Definition of exponents} 
Several studies (see Table 2)  have characterized  the yielding transition with several exponents, which we now recall.

{\it Flow curves:} Rheological properties are singular near the yielding transition. Herschel and Bulkley noticed \cite{herschel1926} that for many yield stress materials,   $\Sigma=\Sigma_c+A\dot\gamma^n$, where $\dot{\gamma}$ is the macroscopic strain rate and $\Sigma$ is the external shear stress. 
By analogy with depinning we instead introduce the exponent $\beta=1/n$, such that
\be
\label{1}
\dot{\gamma} \sim (\Sigma-\Sigma_c)^{\beta}.
\ee
In contrast to depinning, one finds $\beta>1$ in the yielding transition, as we explain below. Our analysis below focuses on the regime $(\Sigma-\Sigma_c)/\Sigma_c\ll 1$; effects not discussed here are expected to affect the flow curves at larger stresses\cite{Bonnecaze2010, Olsson2007}.

{\it Length scales:} near the yielding transition the dynamics becomes more and more cooperative,  and is correlated on a length scale $\xi$:
\be
\label{2}
\xi \sim |\Sigma-\Sigma_c|^{-\nu}
\ee

{\it Avalanche statistics:}
At threshold $\Sigma=\Sigma_c$, the dynamics occur by avalanches whose size we define as $S \equiv\Delta\gamma L^d$, where $\Delta \gamma$ is the plastic strain increment due to the avalanche, and $L^d$ is the volume of the system. The normalized avalanche distribution  $\rho(S)$ follows a power-law:
\be
\label{4}
\rho(S)\sim S^{-\tau} 
\ee
In a finite system of size $L$, this distribution is cut-off at some value $S_c$, where the linear extension of the avalanche is of order $L$, enabling to define the fractal dimension $d_f$:
\be
\label{7}
S_c\sim L^{d_f}
\ee

 

A key exponent relates length- and time- scales. $z$ characterizes the duration $T$ of an avalanche whose linear extension is $l$:
\be
\label{5}
T\sim l^{z}
\ee

{\it Density of shear transformations:} If an amorphous solid is cut into small  blocks containing several particles, one can define how much stress $x_i$ needs to be applied to the block  $i$ before an instability occurs.  The probability distribution $P(x)$ is a measure of how many putative shear transformations are present in the sample \cite{Lin2014}. Near the depinning transition, a similar quantity can be defined,  and in that case it is well known that $P(x)\sim x^0$ \cite{Fisher1998}. We have argued \cite{Lin2014} that it must be so when the interaction kernel ${\cal G}$ is {\it monotonic}, i.e. its sign is constant in space. For an elastic interface this is the case, as a region that yields will always destabilize other regions. This implies that locally the distance to  instability $x_i$ always decreases with time, until $x_i<0$ when the block $i$ rearranges. Thus nothing in the dynamics allows the block $i$ to forecast that an instability approaches, and no depletion nor accumulation is expected to occur near $x_i=0$. By contrast for the yielding transition, the sign of ${\cal G}$ varies in space. Thus locally $x_i$ jumps both forward and backward, performing some kind of random walk. Since $x=0$ acts as an absorbing boundary condition (as the site is stabilized by a finite amount once it yields), one expects that  depletion can occur near $x=0$ \cite{Lin2014,lemaitre2007plastic,Hebraud1998}. In \cite{Lin2014} we argued that $P(x)$ must indeed vanish at $x=0$ if the interaction is sufficiently long-range (in particular if $|{\cal G}|\sim 1/r^d$, as is the case for the yielding transition), otherwise the system would be unstable: a small perturbation at the origin would cause extensive rearrangements in the system. 
 Thus the yielding transition is affected by an additional exponent $\theta$ that does not enter the phenomenology of depinning problem:
\be
\label{3}
P(x)\sim x^{\theta}
\ee
with $\theta>0$. Using elastoplastic models we previously measured $\theta\approx0.4$ for $d=3$ and $\theta\approx 0.6$ for $d=2$ \cite{Lin2014}, as we confirm here with improved statistics. 

The definitions of the relevant exponents  are summarized in Table \ref{exponents}, and their values as reported in the literature in Table \ref{table2}.

\begin{figure}[hbt!]
\label{illustration}
  \centering\includegraphics[width=.427\textwidth]{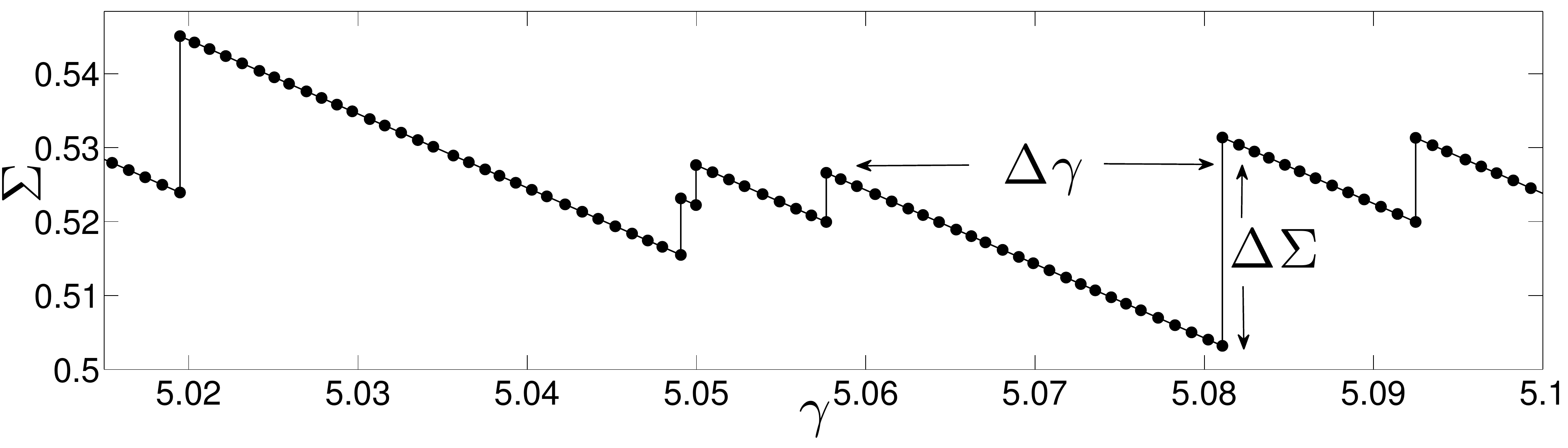} 
  \caption{\small{Example of a stress $\Sigma$ {\it vs}. plastic strain $\gamma$ signal from extremal dynamics simulations of our elasto-plastic model which corresponds to quasi-static strain simulations in computer experiments. Avalanches relax the shear stress by some amount $S/L^{d}$. $\Delta\Sigma$ is the stress increment needed to trigger a new avalanche. In the stationary state, these two quantities must be equal on the average.}}\label{stressstrain}
\end{figure}

\section{Scaling relations}

We now propose several scaling relations, which essentially mirror arguments made in the context of the depinning transition (see supplementary information (S.I.)), with the additional feature that $P(x)$ is singular.

{\it Stationarity:} Consider applying a quasi-static strain in a system of linear size $L$, as represented in Fig.(\ref{stressstrain}). The stress $\Sigma$ fluctuates due to avalanches: 
 an avalanche of size $S$ leads to a stress drop proportional to the plastic strain $\Delta \gamma\sim S/L^d$, of average $\langle S\rangle/L^d$. Using Eqs.(\ref{4},\ref{7}) and assuming $2>\tau>1$
 one gets $\langle S\rangle\sim L^{d_f(2-\tau)}$, so that the average drop of stress is of order $L^{d_f(2-\tau)-d}$.
 
 In between plastic events, the system loads elastic energy, and stress rises  by some typical amount $\Delta \Sigma$. $\Delta \Sigma$ is limited by the next plastic event, and is thus inversely proportional to the rate at which avalanches are triggered. Although one might think that if the system is twice larger, a plastic event will occur twice sooner (implying $\Delta  \Sigma\sim 1/L^d$), this is not the case and $\Delta  \Sigma$ depends on system size with a non-trivial exponent \cite{salerno,Maloney2004,Edan2010,Salerno2013}. As argued in  \cite{Edan2010,lemaitre2007plastic}, one expects $\Delta \Sigma$ to be of order $x_{min}$, the weakest site in the system. If the $x_i$ are independent this implies that $\Delta\Sigma\sim L^{-\frac{d}{\theta+1}}$. In \cite{Edan2010} it was argued based on local considerations that $\theta=1/2$. Instead our recent work \cite{Lin2014} implies that $\theta$ is governed by the elastic interactions between plastic events, and remains a non-trivial exponent that depends on interaction range and spatial dimension. 

Imposing that in a stationary state the average drop and jump of stress must be equal leads to $L^{d_f(2-\tau)-d}\sim\Delta\Sigma$  \cite{salerno,Maloney2004,Edan2010,Salerno2013}; using our estimate of the latter we get $L^{d_f(2-\tau)-d}\sim L^{-\frac{d}{\theta+1}}$,
leading to our first scaling relation:
\begin{equation}
\label{s1}
\tau=2-\frac{\theta}{\theta+1}\frac{d}{d_f}
\end{equation}
As discussed in S.I., a similar but not identical relation holds also for the depinning transition \cite{Fisher1998, Aragon2012}.

{\it Dynamics:}
A powerful idea in the context of the depinning transition is that avalanches below threshold and flow above threshold are intimately related \cite{Fisher1998}. Above threshold,  the motion of the interface 
can be thought as consisting of a number of individual avalanches of spatial extension $\xi$,  acting in parallel. We propose the same image for the yielding transition. If so, the strain rate $\dot \gamma$ in 
the sample is simply equal to the characteristic strain rate of an avalanche of size $\xi$, leading to:
\begin{equation}
\dot{\gamma}=\frac{S}{T\xi^d}\sim(\Sigma-\Sigma_c)^{\nu(d-d_f+z)}.
\end{equation}
implying our second scaling relation, which to our knowledge was not proposed in this context:
\begin{equation}
\beta=\nu(d-d_f+z)\label{beta}
\end{equation}


{\it Statistical tilt symmetry:} If  flow above $\Sigma_c$ consists of independent avalanches of size $\xi$, then the avalanche-induced fluctuations of stress on that lengthscale, $\delta \Sigma$, must be  of order:
\be
\label{12}
\delta \Sigma\sim S_c/\xi^d\sim (\Sigma-\Sigma_c)^{\nu(d-d_f)}
\ee
One expects 
that the fluctuations of stress on the scale $\xi$ must be of order of the distance to threshold $\Sigma-\Sigma_c$. Eq.(\ref{12}) then leads to:
\be
\label{s3}
\nu=\frac{1}{d-d_f}
\ee
It was suggested in \cite{Salerno2013} that Eq.(\ref{s3}) may apply at the yielding transition. A similar relation holds for depinning of an interface if the elasticity is assumed to be linear, a non-trivial assumption underlying Eq.(\ref{12}). In that case it can be derived using the so-called statistical-tilt-symmetry. In S.I., we discuss evidence that linearity applies at the yielding transition, enabling us to use this symmetry to derive Eq.(\ref{s3}).

Overall the scaling relations Eqs.(\ref{s1},\ref{beta},\ref{s3}) allow to express the six exponents we have introduced in terms of three, which we choose to be $\theta,d_f,z$. The corresponding relations are indicated in Table \ref{exponents}.

\begin{table*}[!htb]
\caption{\small{ The critical exponents and their expressions. The third column is the three scaling relations we derive in the text. We compare values measured in our elasto-plastic model both in 2d and 3d, with the predictions from the scaling relations.}}
\begin{tabular}{ |c | c | c | c | c | c |}
\hline
  exponent & expression & relations &  2d measured/prediction & 3d measured/prediction \\ \hline
   $\theta$  & $P(x) \sim x^{\theta}$ &        & 0.57  & 0.35    \\[3pt] 
   $z$       &  $T\sim l^{z}$   &  &  0.57 & 0.65   \\[3pt] 
   $d_f$ & $S_c\sim L^{d_f}$ &  &  1.10 & 1.50  \\ [3pt]
   $\beta$   & $\dot{\gamma} \sim (\Sigma-\Sigma_c)^{\beta}$ & $\beta=1+z/(d-d_f)$   & 1.52/1.62   & 1.38/1.41  \\[3pt] 
   $\tau$    & $\rho(S)\sim S^{-\tau}$    & $\tau=2-\frac{\theta}{\theta+1}\frac{d}{d_f}$ &1.36/1.34 & 1.45/1.48 \\[3pt] 
   $\nu$   & $\xi \sim (|\Sigma-\Sigma_c|)^{-\nu}$ & $\nu=1/(d-d_f)$   & 1.16/1.11  & 0.72/0.67    \\[3pt]   
  \hline
\end{tabular} \label{exponents}
\end{table*}

\vspace{-0.4cm}

\section{Elasto-plastic model}
The phenomenological description proposed above may apply to real materials with inertial or over-damped dynamics, as well as to elasto-plastic models, although the yielding transition in these situations may not lie in the same universality class \cite{Salerno2013,nicolas2014}. In what follows we test our predictions in  elasto-plastic models, implemented as in \cite{Lin2014}, whose details are recalled here. 

We consider square (d=2) and cubic (d=3) lattices of unit lattice size with periodic boundary conditions, where each lattice point $i$ can be viewed as the coarse grained description of a group of particles. It is characterized by a scalar stress $\sigma_{i}$,  a local yield stress $\sigma_{i}^{th}$,  and a strain $\gamma_{i}=\gamma^{el}_{i}+\gamma^{pl}_{i}$. The total stress carried by the system is $\Sigma=\sum_i \sigma_i/L^d$. The elastic strain satisfies $\gamma^{el}_{i}\propto\sigma_i$. The plastic strain is constant in time except when site $i$ becomes plastic, which occurs at a rate $1/\tau_c$ if the site is unstable, defined here as $x_i\equiv \sigma_{i}^{th}-\sigma_{i}<0$. For simplicity we consider that $\sigma_{i}^{th}$ does not vary in space, and use it to define our unit stress $\sigma^{th}=1$. $\tau_c$ is the only time scale in the problem, and defines our unit of time.

 When plasticity occurs, the  plastic strain increases locally and the stress is reduced by the same amount $\delta \gamma^{pl}_i=-\delta \sigma_i =\delta x_i$. We assume that $\delta \gamma^{pl}_i=\sigma_i+\epsilon$ where $\epsilon$ is some random number, taken to be uniformly distributed between $-0.1$ and $0.1$. $\epsilon=0$ would correspond to imposing zero local stress after a plastic event (a choice that we avoid as it sometimes leads to periodic dynamics). When a site relaxes it affects the stress level on other sites immediately, such that:
 \be
 \delta x_j= -{\cal G}({\vec r}_{ij})\delta x_i
 \ee
with ${\cal G}({\vec r}_{ij})\propto \cos (4\phi)/r^2$ in an infinite two-dimensional system under simple shear, and where $\phi$ is the angle between the shear direction and ${\vec r}_{ij}$ \cite{Picard2004}.  In a finite system ${\cal G}$ depends on the boundary conditions \cite{Picard2004}. At fixed stress, by definition  ${\cal G}(0)=-1$ and stress conservation implies that the sum of ${\cal G}$ on any line or column of the lattice is zero. At fixed global strain however, one plastic event reduces the stress by  $1/L^d$. When desired, we model this effect by modifying the interaction kernel as follows: ${\cal G}({\vec r}_{ij})\rightarrow {\cal G}({\vec r}_{ij})-1/L^d$.

 In our model the average plastic strain is defined as $\gamma=\frac{1}{L^d}\sum_{i} \gamma^{pl}_{i}$, and the strain rate simply follows $\dot\gamma =\sum_i \langle \delta\dot \gamma^{pl}_i\rangle /L^d=\sum \sigma_{i} \Theta(\sigma_{i}-1)/(\tau_c L^d)$, where $\Theta(x)$ is the heaviside function. Above $\Sigma_c$, the system will  reach a steady state with a finite $\dot\gamma$. Below or in the vicinity of $\Sigma_c$ however,  the system can spontaneously stop. When this happens, to generate a new  avalanche, we trigger the dynamics by giving very small random kicks to the system (chosen to conserve stress on every line and column) until one site becomes unstable.
 
This elasto-plastic model is essentially identical to the automaton models  introduced in  \cite{narayan1994} in the context of the depinning transition, where the role of the plastic strain $\gamma^{pl}_i$ is played by the transverse displacement of the elastic interface $u_i$. The only qualitative difference is the form of ${\cal G}$. 

\begin{figure}[htb!]
 \centering \begin{tabular}{@{}cc@{}}
    \includegraphics[width=.45\textwidth]{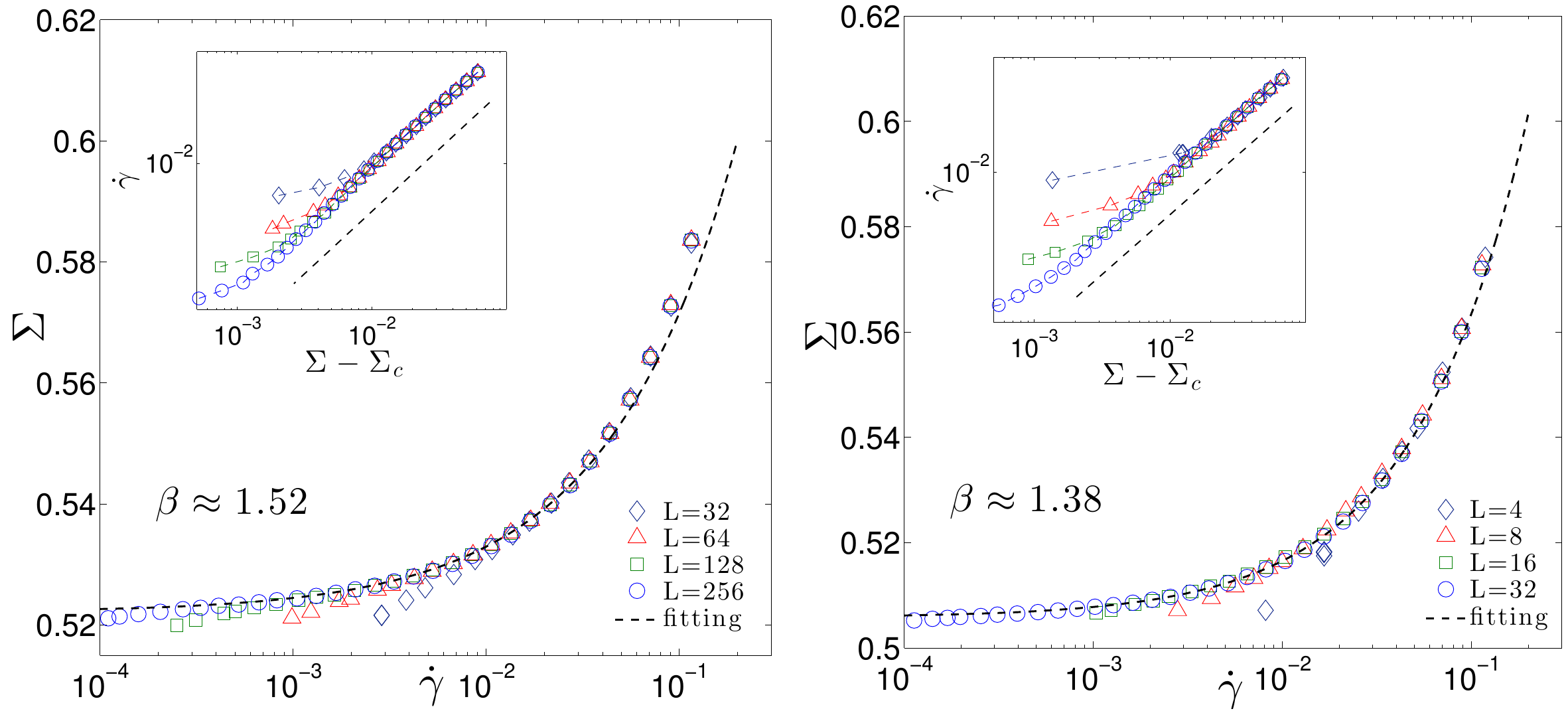} &
    \end{tabular}
  \caption{\small{Insets: flow curves $\dot\gamma(\Sigma)$  in the vicinity of $\Sigma_c$ for different system size $L$ as indicated in legend in $d=2$ (left) and $d=3$ (right). Main curves: the same flow curves in log-linear scale, fitted by the Herschel-Bulkley law, $\Sigma=\Sigma_c+A\dot{\gamma}^{1/\beta}$, which give us $\beta\approx 1.52$ in $2d$, $\beta\approx 1.38$ in $3d$.}}\label{strainratefinitesize}
\end{figure}

\section{Numerical estimation of critical exponents}
{\it Flow curves and length scales}: We first implement the extremal dynamics  protocol: the average stress decreases by $1/L^{d}$ after each plastic event during avalanches, and increases again to generate a new active site at the beginning of a new avalanche. The corresponding stress-plastic strain curves shown in Fig.(\ref{stressstrain}) allows us to estimate the critical stress $\Sigma_c$ and the correlation length exponent $\nu$ from the fluctuations $\delta\Sigma$ of $\Sigma$  at different sizes:
\begin{eqnarray}
&\langle \Sigma_c (L)\rangle = \Sigma_c+ k_1 L^{-\frac{1}{\nu}} +\ldots \nonumber \\
&\delta \Sigma (L) = k_2 L^{-\frac{1}{\nu}} +\ldots
\end{eqnarray}
where $ \langle \Sigma_c (L)\rangle$ is the mean stress and $ \delta \Sigma (L)$ the standard deviation at a given size $L$, and $k_1, k_2$ are non universal constants. From our data (see S.I.), we obtain $\Sigma_c = 0.5221\pm 0.0001$, $\nu=1.16\pm0.04$ for d=2 and $\Sigma_c = 0.5058\pm 0.0002$, $\nu=0.72\pm0.04$ for d=3. These quantities can also be reliably extracted from finite strain rate measurements, as shown in S.I.

We then compute the flow curve at fixed strain rate. The stress is adjusted in order to keep the fraction of unstable sites fixed. The determination of the exponent $\beta$ is very sensitive to the value of $\Sigma_c$. Using the values obtained from the previous analysis we find $\beta=1.52 \pm 0.05$ for $d=2$ and $\beta=1.38 \pm 0.03$ for $d=3$, as shown in Fig.(\ref{strainratefinitesize}).

%

\begin{figure}[hbt!]
 \centering\begin{tabular}{@{}cc@{}}
    \includegraphics[width=.45\textwidth]{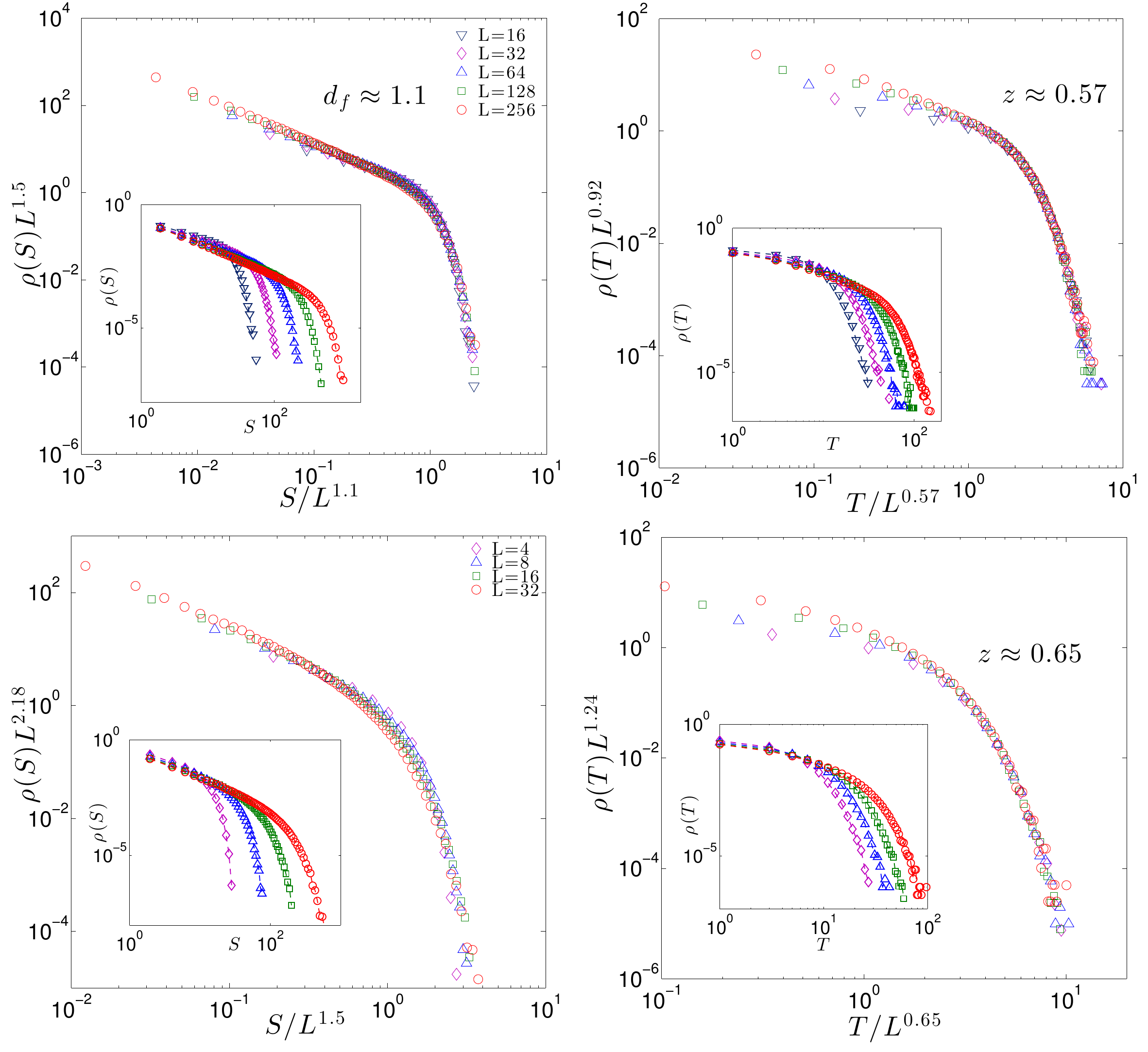} & 
  \end{tabular}
  \caption{\small{Left, Insets: avalanche size distribution $\rho(S,L)$ for extremal dynamics as the system size $L$ is varied in $d=2$ (upper curve) and $d=3$ (bottom curve). Main plots: rescaling avalanche size enable to collapse these distribution, allowing to extract a fractal dimension $d_f$. Right, Insets:  distribution $\rho(T)$ of the duration of the avalanches for the system sizes as indicated in legend in $d=2$ (upper curve) and $d=3$ (bottom curve). Main plots: The cut-off  present in these distribution can be collapsed by rescaling time, leading to an estimate of the dynamical exponent $z$.}}\label{avalanche2}
\end{figure}

{\it Avalanche statistics:} 
Avalanche statistics can be investigated using extremal dynamics and  Fig.(\ref{avalanche2}). As documented in S.I. this method leads for our largest system size to $\tau\approx1.2$ for $d=2$ and $\tau\approx1.3$ for $d=3$. This measure appears to have large finite size effects however. We find that such effects are diminished if we work instead at constant stress $\Sigma$, and consider $\rho(S,\Sigma)$ as $\Sigma\rightarrow\Sigma_c^{-}$. In S.I., we find using this method that $\tau=1.36\pm0.03$ for $d=2$, and $\tau= 1.45 \pm0.05$ for $d=3$.


Next we evaluate the fractal dimension $d_f$ and the dynamical exponent $z$ using extremal dynamics. Here the avalanche cut-off $S_c$ corresponds to avalanche of linear extension $\sim L$, so that for large systems one expects $\rho(S,L)\sim S^{-\tau}h(S/L^{d_f})\sim L^{-d_f\tau}H(S/L^{d_f})$, where $h$ is some function and $H(x)=x^{-\tau}h(x)$. This collapse is checked in Fig.(\ref{avalanche2}) and leads to $d_f=1.10\pm0.04$ for  $d=2$ and $d_f=1.50\pm 0.05$ for $d=3$ (for the collapse we used the values of $\tau$ measured with the constant stress protocol). Error bars are estimated by considering the range of exponents  for which the collapse is  satisfactory. To measure $z$ we record  the duration $T$ of each avalanche, and compute the duration distribution $\rho(T)$ for different system size. These distributions are cut-off at some $T_c$, corresponding to the duration of avalanches of spatial extension $L$, so that $T_c\sim L^z$. As shown in the right panels of Fig.(\ref{avalanche2}) we indeed find a good collapse $\rho(T,L)\sim T^{-\tau^{\prime}}h_2(T/L^z)\sim L^{-\tau^{\prime}z} H_2(T/L^z)$ with  $z=0.57\pm0.03$, $\tau^{\prime}\approx 1.6$  for $d=2$ and $z=0.65\pm0.05$, $\tau^{\prime}\approx 1.9$ for $d=3$.

%

\begin{figure}[hbt!]
\begin{tabular}{@{}cc@{}}
    \includegraphics[width=.46\textwidth]{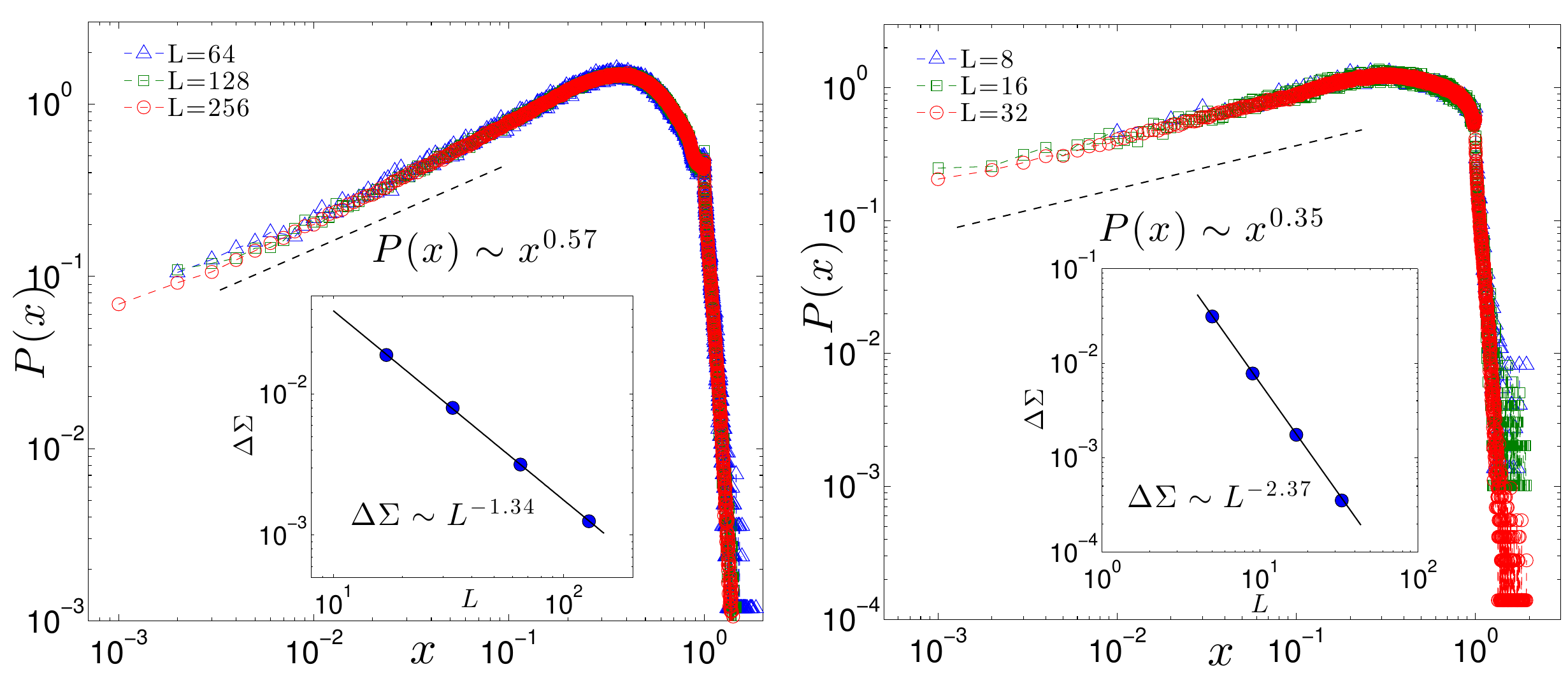} &
    \end{tabular}
  \caption{ Shear transformation distribution $P(x)$ where $x$ is the local distance to an instability for $d=2$ (left) and $d=3$ (right). Insets: amplitude of stress increments $\Delta \Sigma$ before an instability occurs as a function of $L$, found to follow $\Delta \Sigma\sim L^{-d/(1+\theta)}$. }\label{quadtheta}
\end{figure}

 {\it Density of shear transformations:}
In elastoplastic models it is straightforward to access the  local distance to thresholds $x_{i}$ and to compute its distribution $P(x)$ \cite{Lin2014}. Here we recall these results with improved statistics. We fix the stress at $\Sigma_c$, and let the system evolve for a long enough time such that $\gamma\gg1$. The dynamics occasionally stops; at that point we measure $P(x)$, and average over many realizations. 
As shown in Fig.(\ref{quadtheta}), we find $\theta=0.57\pm0.01$ for $d=2$, $\theta=0.35\pm0.01$ for $d=3$, where the error bar is from the error estimation of linear fit. Although in experiments $P(x)$ is hard to access, the system size dependence of the average increment of stress where no plasticity occurs should be accessible, and follows $\Delta \Sigma\sim L^{-d/(1+\theta)}$. In the insets of  Fig.(\ref{quadtheta}) $\Delta \Sigma$ is computed via extremal dynamics, leading to slightly smaller exponents $\theta\simeq 0.50$ for $d=2$ and $\theta\simeq0.28$ for $d=3$, a difference presumably resulting from corrections to scaling.

\subsection{Theory {\it \normalsize{vs}} numerics} 
\ Our scaling relations can now be tested, and this comparison is shown in Table \ref{exponents}. We find very good agreements for all the three scaling relations, Eq.(\ref{s1},\ref{beta},\ref{s3}).


\begin{minipage}[!hbt]{0.42\textwidth}
\begin{table*}[!hbt]
\caption{\small{Values of exponents as reported in the literature. 
The exponents characterizing the relationships between length and strain rate $\xi\sim \dot{\gamma}^{-\nu/\beta}$, average avalanche size and  system length  $\langle S\rangle\sim L^{(2-\tau)d_f}$ and avalanche durations with sizes $T\sim S^{z/d_f}$ are often reported, and are shown here.  Three-dimensional observations are labeled (3d), otherwise the value correspond to two-dimensional systems.} }\label{table2}

\scalebox{0.8}{\begin{tabular}{ |c | c | c |  l  |  l  |  l | l |}
\hline
  exponent & values(2d) &values(3d) & lattice model & molecular dynamics &   experiments  \\    \hline
   $\beta$  &  1.52  & 1.38 &  1.78\cite{nicolas2014}  & 2\cite{Chaudhuri2012}, 2.33\cite{lerner}, 3(3d)\cite{lerner}  &  2.22(3d)\cite{Cloitre2003}, 2.78\cite{van_Hecke_foam_HB}, 2.22(3d)\cite{emulsion_HB} \\[4pt] 
   $\nu/\beta$  &  0.72  & 0.53 &   0.5\cite{Picard2005}, 0.6\cite{nicolas2014}, 1\cite{Martens2011}  & 0.5\cite{caroli}, 0.43\cite{lerner}, 0.33(3d)\cite{lerner} & \\[4pt] 
   $\tau$  &  1.36  & 1.43  & 1.34\cite{Budrikis2013},1.25\cite{Talamali2011} &1.3 \cite{Salerno2013}, 1.3(3d)\cite{Salerno2013} & 1.37--1.49(3d)\cite{Sun2010}, 1.5(3d)\cite{Dahmen2014}\\[4pt] 
   $d_f$  &  1.1  & 1.5  & 1.5\cite{Martens2011}, 1.5\cite{nicolas2014}, 1\cite{Talamali2011} &  0.9\cite{Salerno2013}, 1.1(3d)\cite{Salerno2013},1\cite{Arevalo2014}, 1.5(3d)\cite{Arevalo2014}, 1.6(3d)\cite{Bailey2007}  &   \\[4pt] 
  $(2-\tau)d_f$  &  0.7 & 0.8  & 0.75\cite{Talamali2011} & 1\cite{Maloney2004}, 0.6\cite{Salerno2013}, 0.8(3d)\cite{Salerno2013} &\\[4pt] 
  $z/d_f$  & 0.52 &  0.43 &   0.68\cite{Budrikis2013} &  &  0.5\cite{Dahmen2014}   \\[4pt]
  $\theta$  &  0.57  & 0.35   &     & 0.54\cite{Salerno2013}, 0.43(3d)\cite{Salerno2013}, 0.5\cite{Edan2010},0.5(3d)\cite{Edan2010}   &\\[4pt] 
  \hline
\end{tabular}}
\end{table*}
\end{minipage}

\vspace{-0.4cm}
\section{
Comparison with MD and experiments}
Although elastoplastic models are well-suited to test theories, they make many simplifications, and thus may not fall in the universality class of real materials. 
One encouraging item is our estimate of $\theta$, which is very similar to the value extracted from finite size effects in MD simulations using overdamped dynamics, as reported in Table.\ref{table2}. This is consistent with our finding \cite{Lin2014} that $\theta$ (and $\tau$) is independent of the choice of dynamical rules in our model, that can however dramatically affect the dynamics. Concerning the latter, our choice that the interaction is instantaneous in time, while still being long-range, is likely to affect the exponents $z$ and $\beta$. We expect that if a more realistic time-dependent interaction kernel ${\cal G}({\vec r},t)$ is considered (a costly choice numerically), the exponent $z$ will satisfy $z\geq 1$. According to Eq.(\ref{beta}), this will lead to larger values of $\beta$, in agreement with experiments.

The scaling relations for $\tau$ and $\nu$ in Table \ref{exponents} appear to be supported by MD simulations. In \cite{Salerno2013} for overdamped dynamics, $d_f=0.9$ and $\theta\approx 0.54$ for $d=2$, whereas $d_f=1.1$ and $\theta\approx 0.43$ for $d=3$, leading to $\tau\approx 1.2$ in both $2d$ and $3d$, which compares  well to their measured value $\tau=1.3\pm0.1$. In $d=2$, all numerics \cite{Salerno2013,Arevalo2014} report $d_f\approx 1$, leading to $\nu\approx 1$ as observed in \cite{caroli}. In $d=3$, there is some disagreement on the value of $d_f$: \cite{Arevalo2014,Bailey2007} report $d_f\approx 1.5$ as we do in our elasto-plastic model, in disagreement with \cite{Salerno2013} for which $d_f<d/2$, implying that $\nu<2/d$. It would be useful to resolve this discrepancy, since in the depinning problem, when $\nu<2/d$ another length scale enters in the scaling description, which affects in particular finite size effects \cite{narayan1994,myers1993}. In this situation however, we expect our scaling description to be unchanged if $\nu$ is meant to characterize the correlations of the dynamics for $\Sigma>\Sigma_c$. 

%
%
%
%


\vspace{-0.4cm}

\section{Conclusion}
We have proposed a scaling description of stationary flow in soft amorphous solids, and it is interesting to reflect if this approach 
can apply to other systems. Plasticity in crystals shares many similarities with that of amorphous solids, and the far-field effect of a moving 
dislocation is essentially  identical to the effect of a shear-transformation \cite{Zaiser2006}. Thus we expect that the stability argument of \cite{Lin2014} on the density of regions about to 
yield also applies in crystals, leading to a non-trivial exponent $\theta$ in that case too. Our scaling relations may thus hold  in crystals, although
 the formation of structures such as domain walls could strongly affect the yielding transition. 

Avalanches of plasticity are seen in granular materials where particles are hard \cite{Amon2012,crassous}. However, we believe that at least for over-damped systems this behavior is only transient, and that the elasto-plastic description does not apply for such materials  under continuous shear. Some of us have argued that in that case, a picture based on geometry applies \cite{Lerner2012,lernerpnas}, which also leads to a diverging length scale, but of a different nature \cite{during14}.

The scaling relations proposed here do not fix the values of the exponents, in particular that of $\theta$. To make progress, it is tempting to seek a mean-field description of this problem, 
that would apply beyond some critical dimension. Current mean-field models in which the interaction is random and does not decay with distance lead to $\theta=1$ \cite{Hebraud1998,Lin2014}.
However, the anisotropy is lost in this view, and the fact that $\theta$ diminishes as $d$ increases when anisotropy is considered suggests that a mean-field model that includes anisotropy is needed.
Such a model would be valuable to build  a hydrodynamic description of flow, that would apply for example for slow flow near walls \cite{pouliquen,goyon2008}, a problem for which current descriptions do not include the role of anisotropy  \cite{kamrin,Bocquet2009}. 

\begin{acknowledgments}
We thank  Eric DeGiuli, Alaa Saade, Le Yan, Gustavo D\"uring, Bruno Andreotti,  Mehdi Omidvar, Stephan Bless, and Magued Iskander for discussions. MW acknowledges support from NYU Poly Seed Fund Grant M8769, NSF CBET Grant 1236378, NSF DMR Grant 1105387, and MRSEC Program of the NSF DMR-0820341 for partial funding.
\end{acknowledgments}

\section*{\LARGE{Supplementary Information}}

\section{Numerical evaluation of the critical stress}

\begin{figure}[hbt!]
  \centering\includegraphics[width=.36\textwidth]{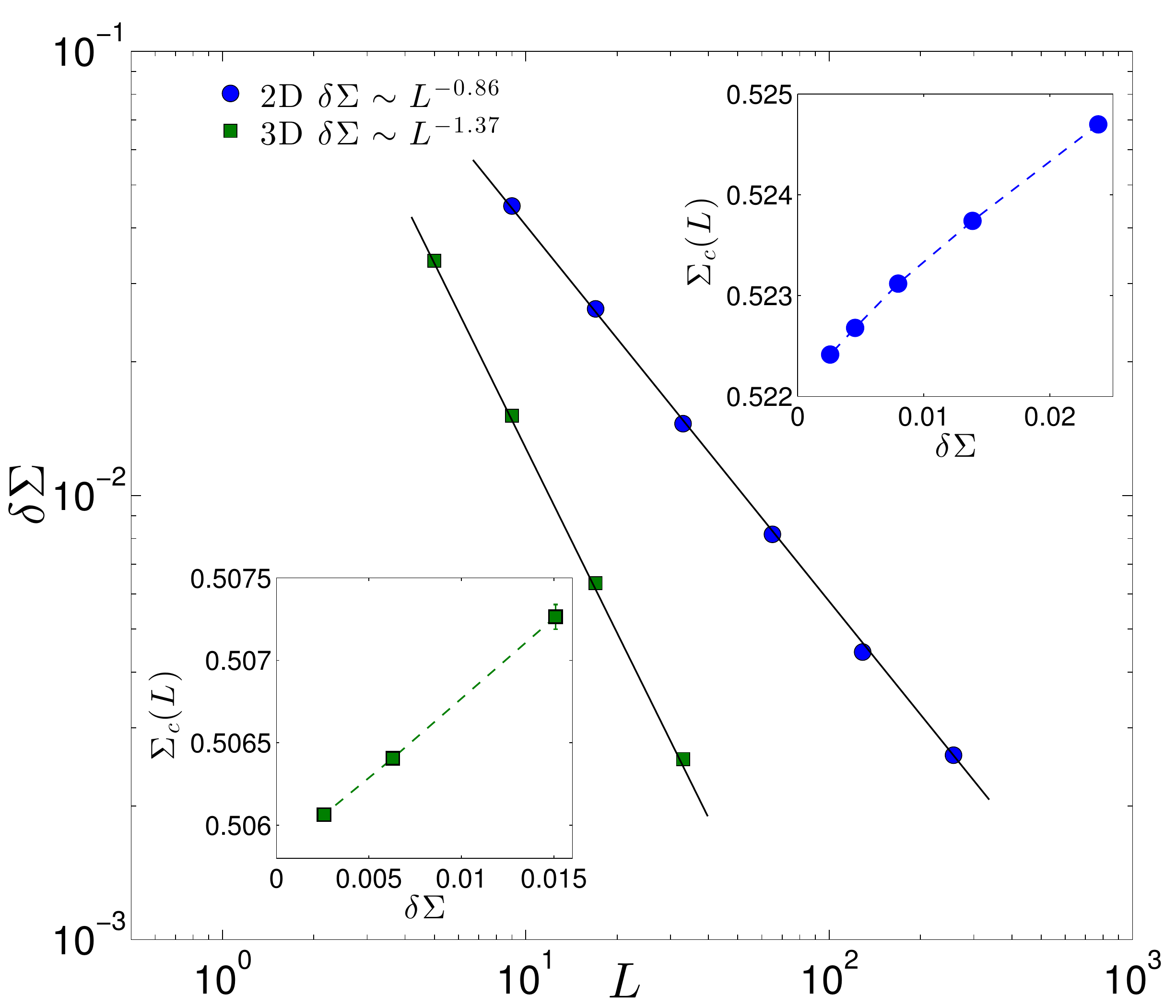} 
  \caption{\small{Stress fluctuations for the extremal dynamics simulation, from which we extract the correlation length exponents  $\nu=1.16\pm 0.04$ in $2d$, and $\nu= 0.72\pm 0.04$ in $3d$. Insets: the relation between between $\Sigma_c(L)$ and $\delta\Sigma$. The intersection of the curves with the $y$ axis yields $\Sigma_c$ in the thermodynamic limit.}}\label{quasistress}
\end{figure}

 Using the extremal dynamics protocol, the system evolves to the critical point  with an average stress $\langle \Sigma_c(L) \rangle$ and stress fluctuations  $\delta\Sigma$  in the stationary state with a dependence of system size as Eq.(13) in the main text. In the insets of Fig.(\ref{quasistress}), we plot out the $\langle\Sigma_c(L)\rangle$ as a function of $\delta\Sigma(L)$, and the critical stress in the thermodynamic limit is just the intersection of the curves with $y$ axis, and we get $\Sigma_c = 0.5221 \pm 0.0001$ for $d=2$ and  $\Sigma_c =0.5058 \pm 0.0002$ for $d=3$. From the dependence of $\delta\Sigma$ on $L$, shown in Fig.(\ref{quasistress}),  we also extract $\nu=1.16\pm 0.04$ in $2d$ and $\nu=0.72\pm 0.04$ in $3d$.

 \section{Fixed stress protocol}
At fixed stress in a finite size system, the dynamics will eventually stop. To trigger a new avalanche we give random kicks to all sites, of amplitude $\delta x_j$, while keeping $\Sigma$ fixed. We consider two methods. In the first one, a site $i$ is chosen randomly, and the amplitude of the kicks follows:
\begin{equation}
\delta x_j= -A{\cal G}({\vec r}_{ij})  \label{111}
\end{equation}
where $A$ is a constant  adjusting the amplitude of kicks. Data presented in the text correspond to $A=1$, but choosing smaller values of $A$ such as $0.1$ did not affect the results, see  Fig.(\ref{avalanchecheck}). Eq.(\ref{111}) ensures that the stress is constant. If no sites become unstable, another site is chosen randomly and another set of kicks following Eq.(\ref{111}) are given. In this method, the site $j_0$ that eventually becomes unstable was typically close to an instability before the random kicks were given. However $j_0$ is not necessarily the weakest site in the entire system.

 In the second method, the dynamics is triggered by imposing that the weakest site $o$ (i.e. $x_o<x_i$ for all $i\neq o$) yields. According to our automaton model this leads to a change of local distance to instability  everywhere in the system, which follows
\begin{equation}
\delta x_j= -{\cal G}({\vec r}_{oj}) \label{112}
\end{equation}
and can lead to avalanches.  We find that these two methods give consistent results for $\tau$, as shown in Fig.(\ref{avalanchecheck}).

\section{Finite size collapse of the flow curve}
\begin{figure}[hbt!]
  \centering\includegraphics[width=.5\textwidth]{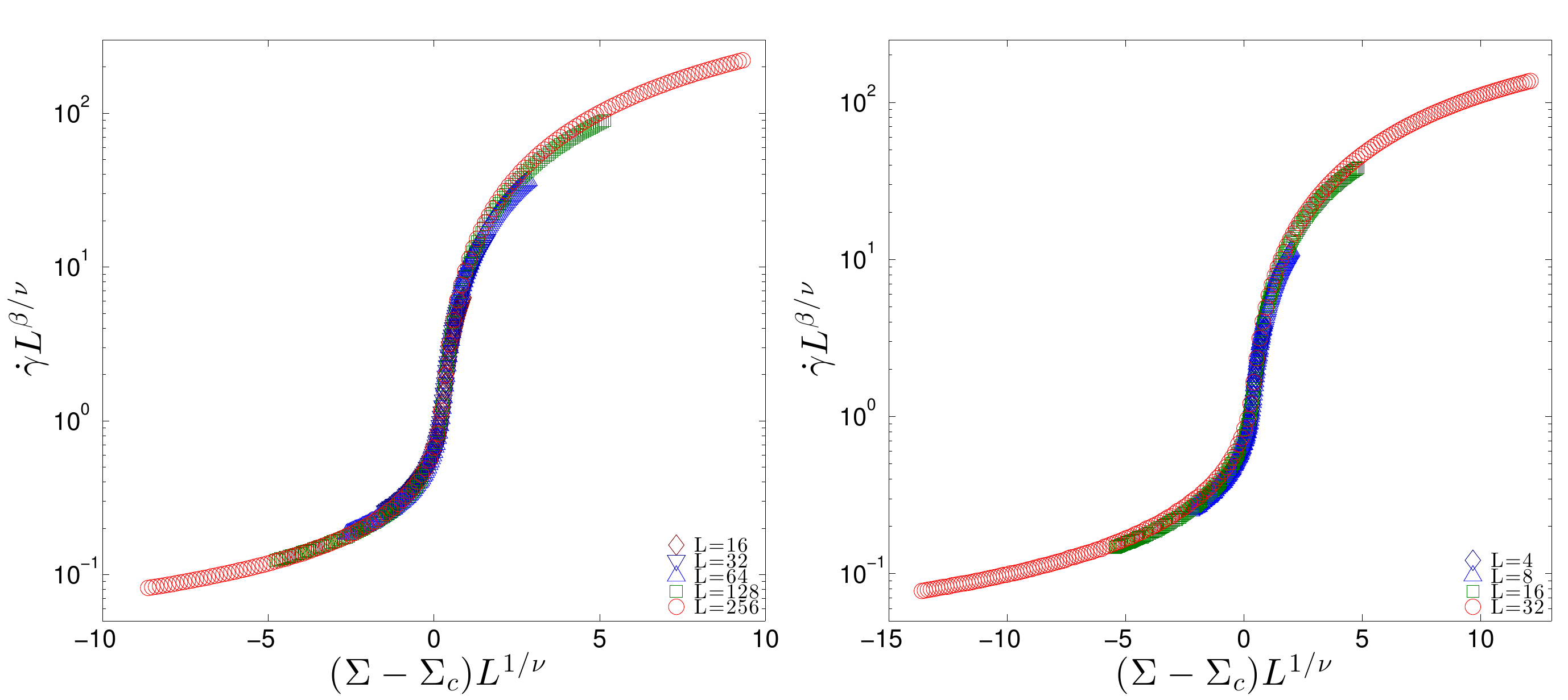} 
  \caption{\small{Collapse of the flow curves obtained from Eq.(\ref{collapse}) using the estimations of $\beta$, $\Sigma_c$ and $\nu$ given in Table 2 of the main text for $d=2$ (Left) and $d=3$  (Right).}}\label{strainratecollapse}
\end{figure}

Our estimations of the threshold $\Sigma_c$ and the correlation length exponent $\nu$ are obtained in the main text using the extremal dynamics protocol. We obtain the same results if we use the fixed stress protocol, with which we can compute  the size-dependent flow curve relating the strain rate, $\dot{\gamma}$, as a function of the external stress, $\Sigma$. From general arguments of finite size scaling, we expect:
\begin{equation}
\dot{\gamma}\sim L^{-\beta/\nu} f((\Sigma-\Sigma_c)L^{-1/\nu})\label{collapse}
\end{equation}
To test the consistency of our methods, in Fig.(\ref{strainratecollapse}) we collapse the different flow curves using Eq.(\ref{collapse}) and the value of $\Sigma_c$, $\nu$ and $\beta$ initially obtained with the extremal dynamics protocol.
We observe a satisfying collapse without any free parameter.

\section{Avalanche statistics}
\begin{figure}[hbt!]
  \centering\includegraphics[width=.48\textwidth]{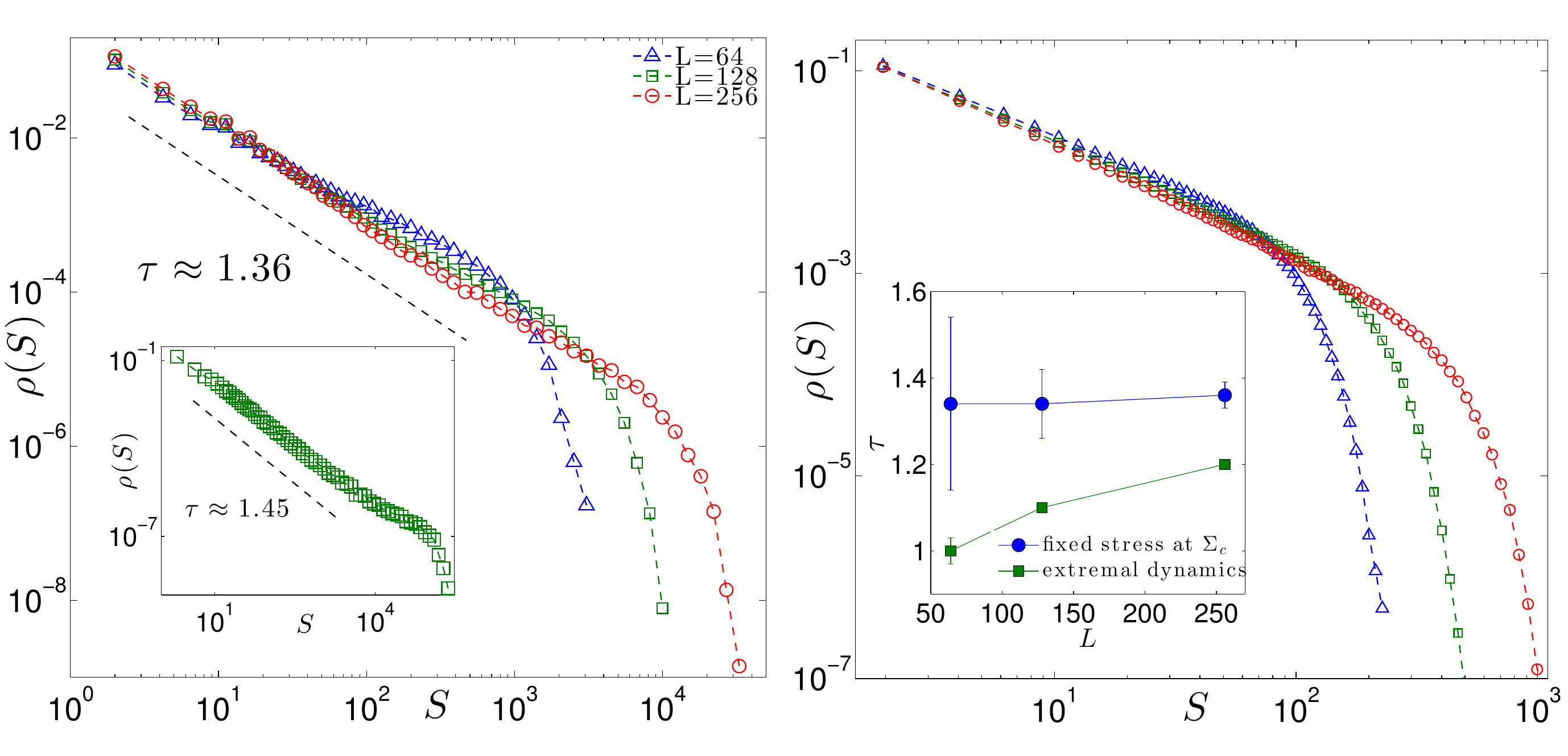} 
  \caption{\small{Left: avalanche distributions at $\Sigma_c$ in $2d$. These distributions clearly overlap for a range of sizes growing with $L$. In this range we obtain an exponent $\tau=1.36\pm0.03$. Here $A=1$,  we check the distributions with a much gentle $A=0.1$, and find the same $\tau$. Inset: avalanche distribution in $3d$ leading to $\tau=1.45\pm0.05$ for $L=64$. Right:
  avalanche distribution obtained using extremal dynamics with the same three sizes. These distributions do not clearly overlap, leading to exponents apprantly increasing with system size. The inset shows how $\tau$ changes with sizes using these two different methods.}}\label{avalanchefinitesize}
\end{figure}

\begin{figure}[hbt!]
  \centering\includegraphics[width=.48\textwidth]{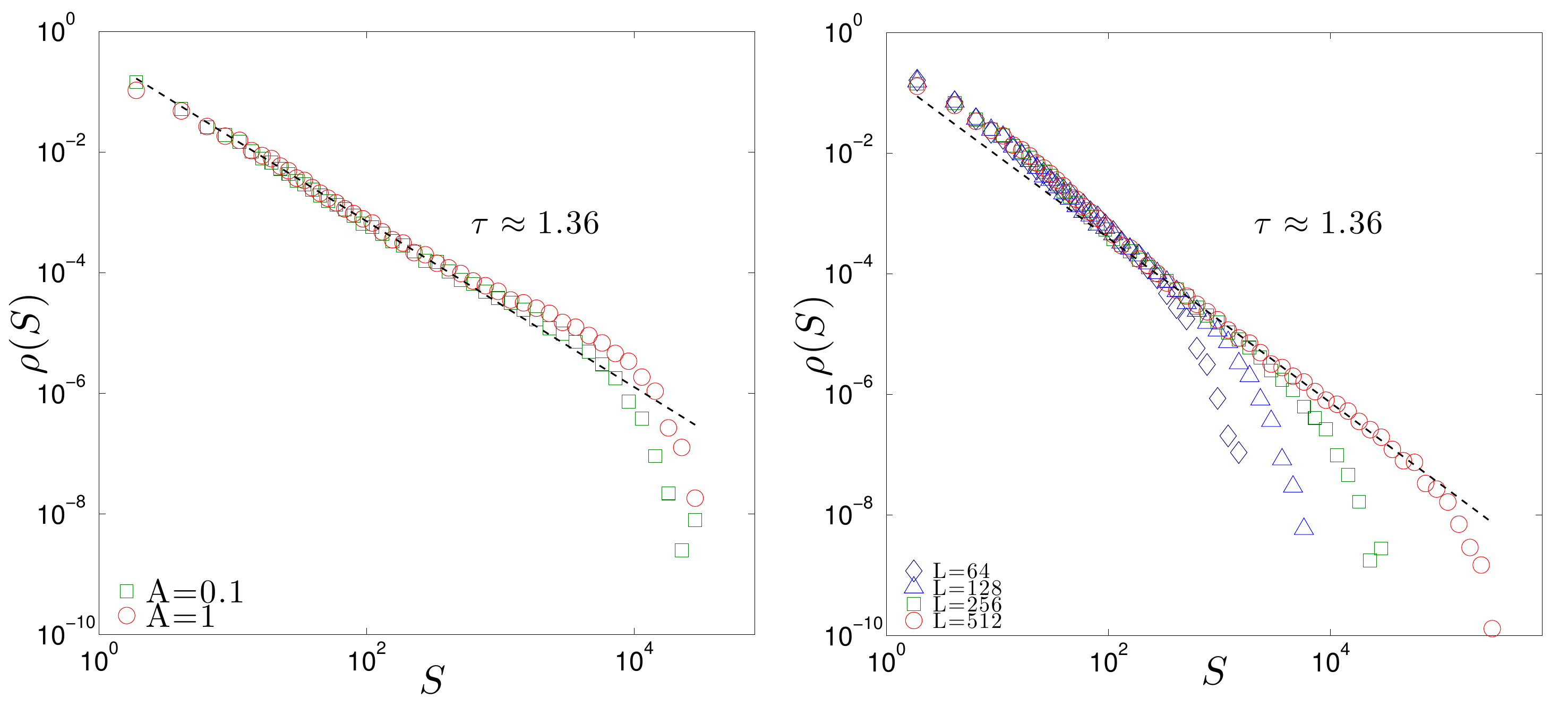} 
  \caption{\small{Left: Effect of the magnitude of random kicks $A$ on the estimation of  $\tau$ for the first method of fixed stress simulation, for $L=256$ and $d=2$.  Right: normalized avalanche probability distribution extracted using the second method  of fixed stress simulation (for which the weakest site yields). This method yields the same exponent $\tau=1.36$.}}\label{avalanchecheck}
\end{figure}

To extract the avalanche distribution exponent $\tau$ accurately, we compare two protocols: (i) constant stress at $\Sigma_c$ and  (ii) extremal dynamics, as shown in Fig.(\ref{avalanchefinitesize}). It turns out that the avalanche distributions in extremal dynamics have  stronger finite size effects than at constant stress. It is thus difficult to extract the avalanche exponent $\tau$ accurately using extremal dynamics. From the inset of Fig.(\ref{avalanchefinitesize})(right), $\tau$ doesn't change significantly with system sizes in  the constant stress method, in contrast to the estimate of $\tau$ that increases with $L$ in extremal dynamics. To extract $\tau$ accurately, we fix the stress at $\Sigma_c$ to collect the avalanche statistics, and we find $\tau=1.36\pm 0.03$ in $2d$, and $\tau=1.45\pm0.05$ in $3d$, and the value of $\tau$ is the same for the two methods of fixed stress protocol, and also insensitive to the value of $A$ in the first method, shown in Fig.(\ref{avalanchecheck}). The error associated to the exponent is estimated by varying the range of avalanche sizes considered in the fit.

\section{General scaling relations}

The three scaling relations derived in the main text for the critical exponents of the yielding transition are similar but not identical to the scaling relations obtained for the depinning transition of an elastic interface.
 In the following we derive three more general relations, namely
 \begin{align}
\nu &= \frac{1}{d-d_f+\alpha_k} \label{STS} \\ 
 \beta &=\nu \left( d-d_f+z\right) \label{beta}\\
 \tau &= 2-  \frac{ d_f-d +1/\nu }{d_f}-  \frac{ \theta}{\theta+1}\frac{d}{d_f} \label{tau}
 \end{align}
 that hold both for yielding and depinning. Here $\alpha_k$ is the dimension of the interaction kernel $\cal {G}$. 
In the context of the yielding transition  $\alpha_k=0$ and $d_f<d$ so that $\beta>1$. 
In the context of the depinning transition $\theta=0$ and $d_f\ge d$, the dimension of the interaction kernel is $\alpha_k=2$ for short range elasticity and  $\alpha_k=1$ for the long range elasticity  of the contact line of a liquid meniscus \cite{joanny84} or of the crack front in brittle materials \cite{gao89, bouchaud11}.

Note that the relations (\ref{beta}) and (\ref{tau}) are expected to be very general, while the first relation is guaranteed only in presence of  statistical-tilt-symmetry, hence only when the interactions are linear. 
For example, it is known that the non-harmonic corrections to the elastic energy can modify the universal behaviour of the depinning transition with critical exponents that  violate  the relation (\ref{STS}) \cite{kardar1998}.
For the yielding transition the validity  of (\ref{STS}) is supported by recent molecular dynamics simulations \cite{SalernoSI2013} that show that the stress decay during an avalanche is proportional to the energy jump, a scaling consistent with linear elasticity. Such linearity is assumed a priori in elasto-plastic models, and is required for the statistical-tilt-symmetry to apply, see below.

\subsection{From the elasto-plastic automaton to the continuum model}
 
The $d$ dimensional  elasto-plastic model studied in this paper is a discrete automaton. Its continuum limit gives the time evolution of the strain field $ \gamma_{\vec r}$ in each point of the space:
\begin{equation}
\label{eq_continuum}
\partial_t \gamma_{\vec r} =  \int_{\vec r'} {\cal G}(\vec r- \vec r') \, \gamma_{\vec r'} +\Sigma +\sigma^{\text{dis}} (\gamma_{\vec r}, \vec r)
\end{equation}
The first term of the equation describes the interactions between the different parts of the system. Note that the interactions are linear in the strain field $\gamma$, and governed by a time-independent interaction kernel, ${\cal G}(\vec r)$. As discussed in the main text, for elastic depinning models the kernel is {\em monotonic}, while for amorphous materials it is {\em non-monotonic}, anisotropic and can be conveniently written in the Fourier space:
\begin{eqnarray}
\label{eq_kernels}
{\cal G} (\vec k) =\left\{
\begin{array}{c}  -\frac{4k_x^2 k_y^2}{k^4},\; \;\; \text{for}\; d=2\\
-\frac{4k_x^2 k_y^2+ k_z^2k^2}{k^4},\; \; \; \text{for}\;d=3. 
\end{array}
\right.
\end{eqnarray}
The other two terms are the external stress, $\Sigma$, and the quenched disorder, $\sigma^{\text{dis}}(\gamma, \vec r)$, which takes into account the inhomogeneities of the local yield stress. In the automaton model the scalar stress $\sigma_i$  corresponds to the sum of the first two terms, and $\sigma^{\text{dis}}(\gamma, \vec r)$ is assumed to be a collection of narrow wells randomly located along $\vec r$. The parameters $\sigma^{th}$, $\epsilon$ and $\tau_c$ are related respectively  to the well depth, to the distance between consecutive wells and to the time needed to move from an unstable well to a stable one.  

Below threshold, $\Sigma<\Sigma_c$, the local strain fields are pinned inside a set of narrow wells. If a small perturbation is applied  (e.g.  a little change in the well locations),  the local strain field responds either (almost everywhere) linearly simply readjusting its value inside the well, either (when a well becomes unstable) with a large modification accompanied by a stress release that can be the seed of a large avalanche. This  non-linear response gives a singular contribution to the susceptibility which becomes important close to $\Sigma_c$. Note that in presence of a  {\em non-monotonic} interaction kernel, the avalanche size $S =\sum_i \Delta \gamma_i$ can be positive or negative, however the positive external stress $\Sigma$ strongly suppress negative avalanches that can, in practice, be neglected.

\subsection{The statistical tilt symmetry}

We now focus on the response of the system when we add to Eq.(\ref{eq_continuum}) a  tilt,  $\sigma^{\text{tilt}}_{\vec r}$, namely an inhomogeneous local stress of zero spatial average.
 In presence of linear interactions, the tilt can be absorbed in a new strain filed $\tilde \gamma_r$ defined as
 \begin{equation}
\tilde \gamma_{\vec r} = \gamma_{\vec r} +  \int_{{\vec r}'}  {\cal G}^{-1}({\vec r} -{\vec r}')  \sigma_{{\vec r}'}^{\text{tilt}}
 \end{equation}
 and governed by the following evolution equation
 \begin{equation}
 \label{tiltless}
\partial_t \tilde  \gamma_r =   \int_{\vec r'} {\cal G}(\vec r- \vec r') \, \tilde \gamma_{\vec r'} +\Sigma +\sigma^{\text{dis}} (\tilde \gamma -   {\cal G}^{-1}  \sigma^{\text{tilt}}, \vec r).
\end{equation}
The latter equation points out that the effect of the tilt can be absorbed with a shift of the location of the narrow wells. Thus, once the average over disorder is taken, the tilt disappears from Eq.(\ref{tiltless}) if the correlation 
$\overline{\sigma^{\text{dis}} ( \gamma , \vec r) \sigma^{\text{dis}} ( \gamma' , \vec r')}$ only depend on $\gamma-\gamma'$. For example in the steady state, when the system becomes independent of the initial conditions,  the average response of  $\gamma_q$ to a tilt $ \sigma^{\text{tilt}}_q$ acting on the mode $q$, is
\begin{equation}
\chi_q=\overline{\frac{\partial  \gamma_q}{\partial \sigma^{\text{tilt}}_q}} = \overline{\frac{\partial  \tilde \gamma_q - {\cal G}_q^{-1} \sigma^{\text{tilt}}_q }{\partial \sigma^{\text{tilt}}_q}} =-{\cal G}_q^{-1} 
\end{equation}
This exact expression should be compared with the scaling behaviour of the singular part of the susceptibility governed by the characteristic scale $\xi \sim (\Sigma_c-\Sigma)^{-\nu}$. In this regime  the strain field grows as $\Delta \gamma \simeq \xi^{d_f-d}$ and  noting that the tilt has the dimension of a stress,  we expect that the singular part of the susceptibility scales as   $\chi^{\text{sing.}}  \sim \xi^{1/\nu+d_f-d}$, which gives  $1/\nu-d+d_f= \alpha_k$, namely Eq.(\ref{STS}).  Here $\alpha_k$ is the dimension of the kernel $1/{\cal G}_q$.  For short range elastic depinning  $\alpha_k=2$, for long range depinning $\alpha_k=1$, while the anisotropic kernel one has $\alpha_k=0$.

\subsection{Stationarity}

Concerning the other two scaling relations: Eq.(\ref{beta}) is identical to the one derived in the main text and Eq.(\ref{tau}) is  still a consequence of  the stationarity of the avalanche dynamics. In general
an avalanche of size S leads to a stress drop which is not simply proportional to the plastic strain, but rather to 
 $\Delta \gamma \,  L^{d-d_f-1/\nu } $ , so that the average stress drop induced by avalanches scales as
\begin{equation}
\Delta \Sigma \sim \frac{\langle S\rangle}{L^d} \frac{L^{-1/\nu}}{L^{d_f -d} }. 
\end{equation}
On the other hand the stress injection before observing a new avalanche scales as $L^{-d/(\theta+1)}$, so that
\begin{equation}
\frac{\langle S\rangle}{L^d} \frac{L^{-1/\nu}}{L^{d_f -d} } \sim  L^{-d/(\theta+1)}.
\end{equation}
Finally, using $\langle S\rangle \sim L^{(2-\tau)d_f}$ we obtain Eq.(\ref{tau}).

\end{article}

\end{document}